\documentclass[a4paper,12pt]{article}

\usepackage[dvipdfmx]{graphicx}
\usepackage{array}
\usepackage{amsmath}
\usepackage{amssymb}
\usepackage{latexsym}
\usepackage{color}

\date{empty}
\begin{document}
\begin{titlepage}
\null
\begin{flushright}
YGHP-20-04 \\
\end{flushright}
\vskip 1.0cm
\begin{center}
{\Large \bf 
Higgs-Portal Dark Matter in Nonlinear MSSM
}
\vspace{1cm}
\normalsize
\renewcommand\thefootnote{\alph{footnote}}

{\large
Masato Arai\footnote{arai(at)sci.kj.yamagata-u.ac.jp} and
Nobuchika Okada $\ddagger$ \footnote{okadan(at)ua.edu}, 
}
\vskip 1.0cm
  {\it
  Faculty of Science, Yamagata University, \\ 
  Yamagata 990-8560, Japan \\
  \vskip 0.2cm
  $^\ddagger$Department of Physics and Astronomy, \\
  University of Alabama, Tuscaloosa, AL 35487, USA
}
\vskip 1.5cm
\begin{abstract}

\vskip 0.5cm
Supersymmetric (SUSY) extension of the Standard Model (SM) is a primary candidate for 
  new physics beyond the SM. 
If SUSY breaking scale is very low, for example, the multi-TeV range,  
  and the SUSY breaking sector, except for the goldstino (gravitino), is decoupled
  from the low energy spectrum, the hidden sector effect in the minimal SUSY SM (MSSM) 
  is well described by employing the goldstino chiral superfield ($X$) with the nilpotent condition of $X^2=0$. 
Although this so-called ``nonlinear MSSM'' (NL-MSSM) provides a variety of interesting phenomenologies, 
  there is a cosmological problem that the lightest superpartner gravitino is too light 
  to be the major component of the dark matter (DM) in our universe. 
To solve this problem, we propose a minimal extension of the NL-MSSM 
  by introducing a parity-odd SM singlet chiral superfield ($\Phi$). 
We show that the interaction of the scalar component in $\Phi$ with the MSSM Higgs doublets 
  is induced after eliminating $F$-component of the goldstino superfield and the lightest real 
  scalar in $\Phi$ plays the role of the Higgs-portal DM. 
With a suitable choice of the model parameters, a successful Higgs-portal DM scenario can be realized.  
In addition, if SUSY breaking scale lies in the multi-TeV range, 
  the SM-like Higgs boson mass of 125 GeV can be achieved by the tree-level Higgs potential 
  through the low-scale SUSY breaking effect. 
\end{abstract}

\end{center}

\end{titlepage}

\newpage
\setcounter{footnote}{0}
\renewcommand\thefootnote{\arabic{footnote}}
\pagenumbering{arabic}
\section{Introduction}
Although the current experimental data show no plausible evidence of new physics beyond the Standard Model (SM), 
  the minimal supersymmetric (SUSY) extension of the SM (MSSM) is still a primary candidate for new physics. 
As has been well-known and intensively studied, 
  the MSSM not only provides us with a solution to the gauge hierarchy problem
  but also offers a variety of interesting phenomenologies, 
  such as the origin of the electroweak symmetry breaking from SUSY breaking, 
  the SM-like Higgs boson mass prediction with soft SUSY breaking parameters, 
  the lightest superpartner (LSP) as a natural candidate for the dark matter (DM) in our universe, 
  and the grand unified theory paradigm with the successful unification of the three SM gauge couplings
  at the scale of ${\cal O}(10^{16}\, {\rm GeV})$. 
Many ongoing and planned experiments will continue searching for the MSSM, or in more general, 
  supersymmetric theories beyond the SM.

In phenomenologically viable models, SUSY is spontaneously broken in the hidden sector  
  and the SUSY breaking effects are mediated to the MSSM sector 
  by a certain mechanism for generating soft SUSY breaking terms in the MSSM. 
Associated with spontaneous SUSY breaking, a massless fermion called goldstino emerges 
  due to the Nambu-Goldstone theorem, and it is absorbed into the spin-$1/2$ component
  of the spin-$3/2$ massive gravitino in supergravity. 
The gravitino mass is characterized by the SUSY breaking order parameter $f$ 
  and the reduced Planck mass of $M_P=2.43 \times 10^{19}$ GeV as $m_{3/2} \simeq f/M_P$.  
It is possible that SUSY breaking occurs at a very low energy 
(see, for example, Ref.~\cite{Brignole:1996fn}). 
If this is the case, gravitino becomes the LSP and is involved in phenomenology at low energies. 
For example, if the SUSY breaking scale lies in the multi-TeV range, 
  the LSP gravitino is extremely light with its mass of ${\cal O}({\rm meV})$. 
Assuming the decoupling of the hidden sector fields except for the light gravitino (or, equivalently, goldstino) 
  the low energy effective theory involving the very light gravitino can be described 
  by employing a goldstino chiral superfield $X$ with the nilpotent condition $X^2=0$ \cite{Rocek:1978nb, Lindstrom:1979kq, Komargodski:2009rz}. 
With this formalism, the phenomenology of the MSSM with the goldstino superfield 
  has been studied in detail \cite{Antoniadis:2010hs, Antoniadis:2012zz, Antoniadis:2014eta} (see also Ref.~\cite{Dudas:2012fa} for the phenomenology in a more general setup). 
This framework is the so-called nonlinear MSSM (NL-MSSM).   
Interestingly, it has been shown that if the SUSY breaking scale lies in the multi-TeV range, 
  the SM-like Higgs boson receives a sizable contribution to its mass at the tree-level 
  after eliminating $F$-component of the goldstino superfield and as a result, 
  the Higgs boson mass of around $125$ GeV can be achieved by the tree-level Higgs potential. 
This is in sharp contrast with the usual MSSM in which the 125 GeV SM-like Higgs boson mass
  is reproduced by quantum corrections through scalar top quarks 
  with the mass larger than multi-TeV.  
In the view point of the collider physics, the NL-MSSM has an advantage that 
  the scalar top quarks can be sufficiently light to be explored in the near future.

The SUSY breaking order parameter $\sqrt{f}\lesssim {\cal O}(100)$ TeV gives the extremely light gravitino 
  with mass $m_{3/2} \lesssim 10$ eV in the NL-MSSM.
Although such a light gravitino is harmless in the phenomenological point of view
  (see, for example, Ref.~\cite{Feng:2010ij}), 
  its relic density is far below the observed dark matter (DM) density. 
Even if the observed relic density is achieved by some non-standard thermal history of the universe, 
  the very light gravitino is likely to be a hot DM and prevents the formation of the observed structure of the universe. 
Therefore, for the completion of the NL-MSSM, we should consider an extension of the model 
  which can supplement the model with a suitable DM candidate. 
In this paper, we propose a minimal extension of the NL-MSSM by introducing a $Z_2$-parity odd 
  SM gauge singlet chiral superfield $\Phi$ and show that the lightest scalar component in $\Phi$ 
  plays the role of the Higgs-portal DM \cite{McDonald:1993ex, Burgess:2000yq}\footnote{For a recent review, see Ref.~\cite{Arcadi:2019lka} and references therein.} through its coupling 
  with the MSSM Higgs doublets induced by the goldstino
 superfield. 
With a suitable choice of the model parameters, we can realize a phenomenologically viable Higgs-portal DM scenario. 
If SUSY breaking scale lies in the multi-TeV range, 
  the SM-like Higgs boson mass of 125 GeV can be achieved by the tree-level Higgs potential 
  through the low-scale SUSY breaking effect.   

\section{NL-MSSM and the Higgs boson mass}\label{Sec:two}
We first present the basic formalism of the NL-MSSM and show 
  how the 125 GeV SM-like Higgs boson mass can be achieved in the framework. 
We begin with the goldstino effective Lagrangian of the form \cite{Komargodski:2009rz}: 
\begin{eqnarray}
{\cal L}_X=\int d^4\theta X^\dagger X +\left(\int d^2\theta fX +{\rm h.c.} \right)\,, \label{eq:Xlag}
\end{eqnarray}
where $X$ is a goldstino chiral superfield, and $f$ is the SUSY breaking order parameter in the hidden sector. 
Although the stability of the hidden sector scalar potential needs an extension of the above minimal K\"ahler potential, 
   this Lagrangian is enough to understand the essence of the formalism.  
The goldstino chiral superfield is subject to the nilpotent condition \cite{Rocek:1978nb, Lindstrom:1979kq, Komargodski:2009rz}, 
\begin{eqnarray}
X^2=0\,.
\end{eqnarray}
which leads us to the expression of
the superfield with the components, 
\begin{eqnarray}
 X={\psi_X \psi_X \over 2 F_X}+\sqrt{2}\theta\psi_X+\theta\theta F_X \,. 
 \label{eq:Xsol}
\end{eqnarray}
The scalar component in the goldstino superfield is to be integrated out in the low energy effective theory, 
  and under the nilpotent condition, it is replaced by the bilinear term of the goldstino fields. 
In fact, substituting Eq.~(\ref{eq:Xsol}) into Eq.~(\ref{eq:Xlag}) and eliminating the auxiliary field $F_X$, 
  we recover the Volkov-Akulov Lagrangian \cite{Volkov:1973ix}.

In the superfield formalism, the spurion technique is a simple way to introduce
  the soft SUSY breaking terms to the MSSM Lagrangian. 
We introduce a dimensionless and SM-singlet spurion field of the form, $Y=\theta^2 m_{\rm soft}$, where
 $m_{\rm soft}$ is a generic notation for the soft terms (denoted $m_{1,2,3}$, $m_{\Psi}$, $m_{\lambda_a}$ in the following),
  and attach it to any SUSY operators in the MSSM.  
The recipe to obtain the NL-MSSM is to replace the spurion by the goldstino superfield as \cite{Komargodski:2009rz}%
\begin{eqnarray}
 Y\rightarrow {m_{\rm soft} \over f}X\,. 
 \label{eq:replace}
\end{eqnarray}
We apply this rule and write the NL-MSSM Lagrangian as follows \cite{Antoniadis:2010hs}: 
\begin{eqnarray}
{\cal L}={\cal L}_0+{\cal L}_X+{\cal L}_H+{\cal L}_m+{\cal L}_{AB}+{\cal L}_g\,. \label{eq:lagNL}
\end{eqnarray}
In the right-hand side, the first term ${\cal L}_0$ denotes the supersymmetric part of the MSSM Lagrangian 
  given by \footnote{For a concise review of the MSSM and the standard notation, see, for example, Ref.~\cite{Martin:1997ns}.}
\begin{eqnarray}
{\cal L}_0&=&\sum_{\Psi, H_u, H_d}\int d^4\theta~\Psi^\dagger e^{V} \Psi \nonumber \\
 &&+\left\{\int d^2\theta~[\mu_H H_dH_u+\lambda_u H_uQU^c+\lambda_d QD^cH_d+\lambda_e LE^cH_d]+{\rm h.c.} \right\} \nonumber \\
 &&+\sum_{\rm a=1}^3{1 \over 4 g_a^2\kappa}\int d^2\theta~{\rm Tr}[W_a^\alpha W_{a\alpha}]+{\rm h.c.}\,,
 \label{eq:SUSY_L}
\end{eqnarray}
where $\Psi=Q, U^c, D^c, L, E^c$, the index $a=1,2,3$ denotes the the SM gauge groups $SU(3)$, $SU(2)$ and $U(1)$, 
  $g_a$ is the corresponding gauge couplings, and $\kappa=1$ for $U(1)$ and $1/2$ for $SU(3)$ and $SU(2)$. 
The vector superfield $V$ in the K\"ahler potential for the chiral superfields implies, for example, 
  $V=2V_3+2V_2+ {1 \over 3} V_1$ for $Q$ etc., where $V_a$ $(a=1,2,3)$ denote the vector superfields of the corresponding SM gauge groups.
${\cal L}_X$ is the hidden sector Lagrangian already introduced in Eq.~(\ref{eq:Xlag}). 
${\cal L}_H$ is the Higgs sector Lagrangian involving the goldstino sueprfield: 
\begin{eqnarray}
{\cal L}_H=-{m_1^2 \over f^2}\int d^4\theta \left( X^\dagger X \right) H_d^\dagger e^{V}H_d
-{m_2^2 \over f^2}\int d^4\theta \left(X^\dagger X \right) H_u^\dagger e^{V}H_u\,.
\end{eqnarray}
The matter field Lagrangian involving the goldstino superfield is given by
\begin{eqnarray}
{\cal L}_m= - \sum_\Psi \left(m_\Psi^2 \over f^2\right)\int d^4\theta \left(X^\dagger X \right) \Psi^\dagger e^V \Psi\,. 
\end{eqnarray}
The bilinear and trilinear SUSY breaking couplings are given by ${\cal L}_{AB}$: 
\begin{eqnarray}
 {\cal L}_{AB}&=&{m_3^2 \over f}\int d^2\theta~XH_dH_u + {\rm h.c.} \nonumber \\
 &&+\int d^2\theta \, X 
  \left\{ \lambda_u \left( \frac{A_u}{f} \right)   U^c H_u Q + 
   \lambda_d \left( \frac{A_d}{f} \right) D^c H_d Q+ 
   \lambda_e  \left( \frac{A_e}{f} \right) E^c H_d L \right \} +{\rm h.c.}
\end{eqnarray}
The last term ${\cal L}_g$ denotes the gauge sector Lagrangian given by 
\begin{eqnarray}
 {\cal L}_g=\sum_{a=1}^3{1 \over 4 g_a^2\kappa}{2 m_{\lambda_a} \over f}\int d^2\theta~X{\rm Tr}[W^\alpha_a W_{a\alpha}]+{\rm h.c.}
\end{eqnarray}

We focus on the Higgs potential in the NL-MSSM, 
 which is read off from ${\cal L}_0+{\cal L}_X+{\cal L}_H+{\cal L}_{AB}$: 
\begin{eqnarray}
V=V_{\rm SUSY} + V_{\rm soft}\,, 
\label{eq:potMSSM}
\end{eqnarray}
where
\begin{eqnarray}
V_{\rm SUSY}&=&
\mu_H^2(|H_u|^2+|H_d|^2) + 
{g_Z^2 \over 8}(|H_u|^2-|H_d|^2)^2+{g_2^2 \over 2}|H_u^\dagger H_d|^2\,, \\
 V_{\rm soft}&=&{\left|f+{m_3^2 \over f}H_uH_d\right|^2 \over 1-{m_1^2 \over f^2}|H_d|^2-{m_2^2 \over f^2}|H_u|^2}\,,  \label{eq:soft}
 \end{eqnarray}
with $g_Z^2 \equiv g_1^2+g_2^2$.  
We express the up-type Higgs and down-type Higgs doublets as 
\begin{eqnarray}
H_u&=&
\begin{pmatrix}
H^+ \\
{1\over \sqrt{2}}(v_u+R_u+i I_u)
\end{pmatrix}  \, , \label{eq:expand1} \\
H_d&=&
\begin{pmatrix}
{1\over \sqrt{2}}(v_d+R_d+i I_d) \\
H^- \\
\end{pmatrix}\,, \label{eq:expand2}
\end{eqnarray}
where $v_u=v\sin\beta$, $v_d=v\cos\beta$ with $v=246$ GeV,  $H^{\pm}$ are charged Higgs fields, 
  and $R_u, I_u, R_d, I_d$ are real scalar fields. 
Substituting them into the Higgs potential, we derive the stationary conditions: 
 \begin{eqnarray}
 {\partial V \over \partial R_u}{\Big |}_0&=&{v \over 4}\left\{
 4\mu_H^2\sin\beta-2M_Z^2 \cos2\beta\sin\beta
 -{2{\cal A} m_3^2\cos\beta \over {\cal B}}+{ {\cal A}^2 m_2^2\sin\beta  \over {\cal B}^2}
 \right\}=0\,, \label{eq:st1} \\
 {\partial V \over \partial R_d}{\Big |}_0&=&{v \over 4}\left\{
 4\mu_H^2\cos\beta+M_Z^2(\cos\beta+\cos 3\beta)
 -{2 {\cal A} m_3^2\sin\beta \over {\cal B}}+{{\cal A}^2 m_1^2\cos\beta  \over {\cal B}^2}
 \right\}=0\,, \label{eq:st2}
\end{eqnarray}
 where ${|_0}$ means that all the fields are taken to be zero, and
\begin{eqnarray}
M_Z^2 &= &  \frac{1}{4} g_Z^2 v^2 \, , \nonumber \\
{\cal A} & = & 4 f^2+m_3^2 v^2\sin 2\beta  \, , \nonumber \\
{\cal B} &=& -2f^2+m_1^2 v^2 \cos^2\beta+m_2^2v^2\sin^2\beta \, . 
\end{eqnarray}
The other stationary conditions such as ${\partial V \over \partial I_u}{\Big |}_0$ are automatically satisfied.
The mass matrix of the CP-even Higgs bosons is given by
\begin{eqnarray}
{\cal M}_{\rm CP-even}=\begin{pmatrix}
{\partial^2 V \over \partial R_d^2}{\big |}_0 & {\partial^2 V \over \partial R_d \partial R_u}{\big |}_0 \\
{\partial^2 V \over \partial R_u \partial R_d}{\big |}_0 & {\partial^2 V \over \partial R_u^2}{\big |}_0\,
\end{pmatrix}\,, \label{eq:mass1}
\end{eqnarray}
while the mass matrices for the CP-odd Higgs bosons and the charged Higgs bosons are
\begin{eqnarray}
{\cal M}_{\rm CP-odd}=\begin{pmatrix}
{\partial^2 V \over \partial I_d^2} {\big |}_0 & {\partial^2 V \over \partial I_d \partial I_u}{\big |}_0 \\
{\partial^2 V \over \partial I_u \partial I_d}{\big |}_0 & {\partial^2 V \over \partial I_u^2}{\big |}_0 \\
\end{pmatrix}\,,~~~~
{\cal M}_{\rm charged}=\begin{pmatrix}
{\partial^2 V \over \partial H^- \partial H^{-*}}{\big |}_0 & {\partial^2 V \over \partial H^- \partial H^+}{\big |}_0 \\
{\partial^2 V \over \partial H^{-*} \partial H^{+*}}{\big |}_0 & {\partial^2 V \over \partial H^{+}\partial H^{+*}}{\big |}_0
\end{pmatrix}\,. \label{eq:mass2}
\end{eqnarray}
%

\begin{figure}[tb]
\centering
\begin{minipage}{0.48\hsize}
\includegraphics[scale=0.6, angle=0]{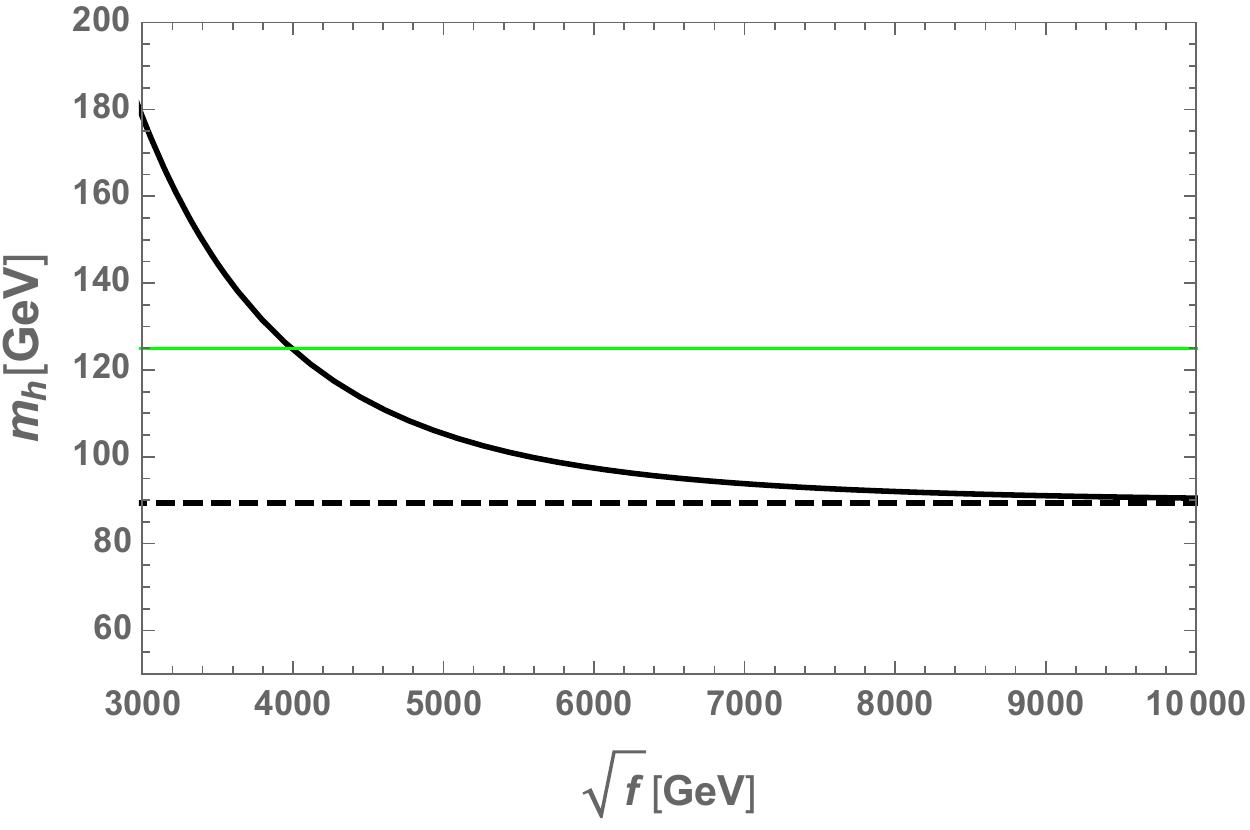}
 \caption{
The SM-like Higgs boson mass ($m_h$) at the tree-level as a function of $\sqrt{f}$ (solid line), 
  along with the standard MSSM prediction at the tree-level (dashed line) 
  and the (green) horizontal line indicting $m_h=125$ GeV. 
In this plot, we have taken $m_1^2=1000^2$ GeV$^2$, $m_2^2=- (2005)^2$ GeV$^2$ and $\tan\beta=10$.
}
\label{fig:SM-Higgs}
\end{minipage}
\hspace{3mm}
\begin{minipage}{0.48\hsize}
\includegraphics[scale=0.6, angle=0]{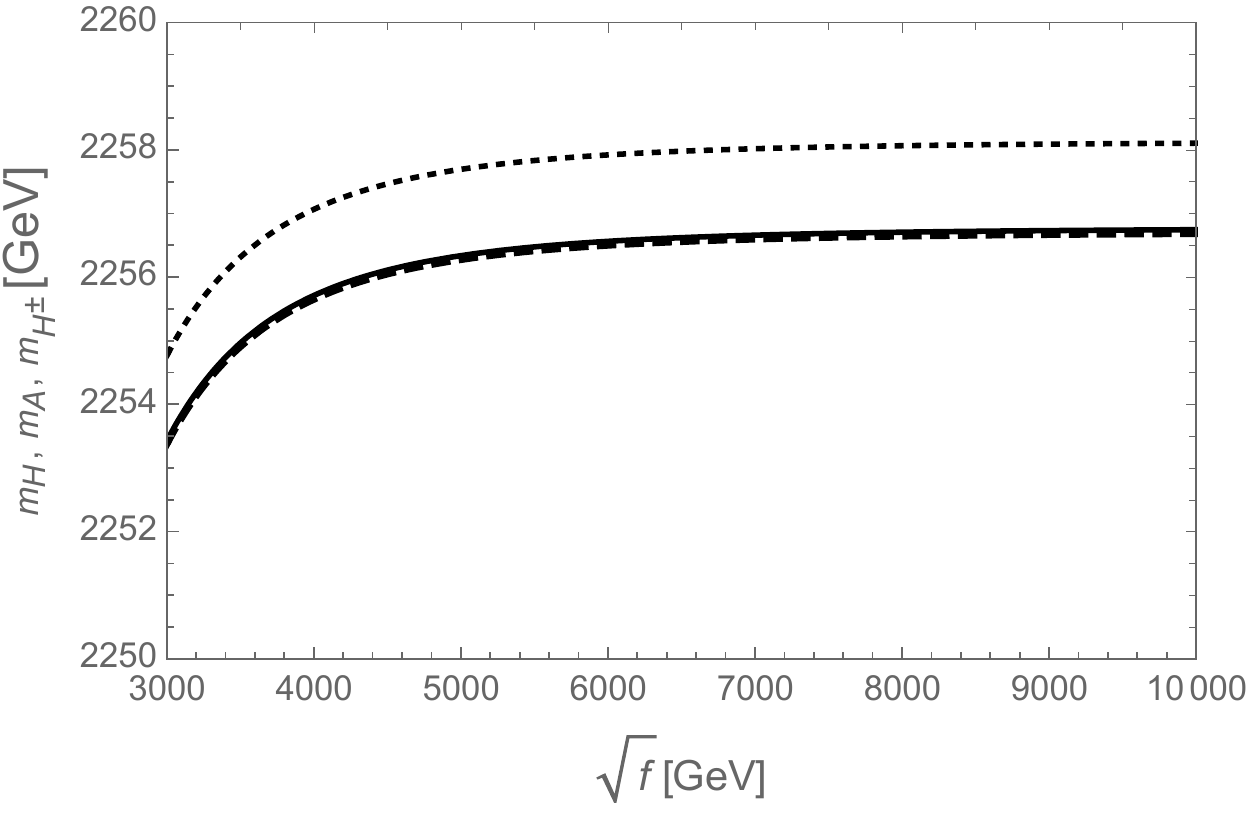}
 \caption{
Same as Fig.~\ref{fig:SM-Higgs} but for the CP-even heavy Higgs boson mass ($m_H$) (solid line),  
   the CP-odd Higgs boson mass ($m_A$) (dashed line) 
   and the charged Higgs boson mass ($m_{H^\pm}$) (dotted line). 
}
 \vspace{4.5mm}
\label{fig:Heavy-Higgs}
\end{minipage}
\end{figure}

By using the above formulas, we numerically calculate the Higgs boson mass spectra. 
First we choose appropriate values for $m_1$, $m_2$, $\tan \beta$ and $\sqrt{f}$ 
  as the input parameters and solve the stationary conditions of Eqs.~(\ref{eq:st1}) and (\ref{eq:st2}) 
  to fix the values of  $\mu_H$ and $m_3^2$. 
We then substitute them into the Higgs potential and calculate the Higgs boson mass eigenvalues 
  from Eqs.~(\ref{eq:mass1}) and (\ref{eq:mass2}). 
Our results are shown in Figs.~\ref{fig:SM-Higgs} and \ref{fig:Heavy-Higgs}.
The solid line in Fig.~\ref{fig:SM-Higgs} shows the mass of the SM-like Higgs boson ($m_h$) 
  as a function of $\sqrt{f}$, where we have fixed $m_1^2=1000^2$ GeV$^2$, $m_2^2=- (2005)^2$ GeV$^2$ and $\tan\beta=10$. 
As $\sqrt{f}$ decreases, the SM-like Higgs boson mass increases 
  from the standard MSSM prediction at the three level $m_h \simeq M_Z\cos 2\beta$ (dashed line) 
  in the limit of $\sqrt{f} \to \infty$. 
The (green) horizontal line indicates $m_h=125$ GeV. 
We find that the main contribution for increasing the SM-like Higgs boson mass comes 
  from the quartic coupling $(m_2^2 |H_u|^2/f)^2$ 
  in the series of expansion of Eq.~(\ref{eq:soft}) and 
  the resultant Higgs boson mass is approximately expressed as
\begin{eqnarray}
 m_h^2 \simeq M_Z^2\cos 2\beta+ 2 \left({m_2^2 \over f}\right)^2 v^2\sin^2\beta\,.
\end{eqnarray}
Therefore, if the SUSY breaking scale is low enough, the SM-like Higgs boson mass of 125 GeV 
  is achieved by the Higgs potential at the tree-level.  
As shown in Fig.~\ref{fig:SM-Higgs}, we have obtained $m_h=125$ GeV for $\sqrt{f}=3990$ GeV.  
If the SUSY breaking scale is larger, the hidden sector effect on the SM-like Higgs boson mass 
 is negligible, so that quantum corrections through heavy scalar top quarks play the crucial role
 to reproduce  $m_h=125$ GeV, as usual in the MSSM.
Fig.~\ref{fig:Heavy-Higgs} shows the masses of the heavy neutral Higgs and the charged Higgs bosons 
  as a function of $\sqrt{f}$ with the same inputs as in Fig.~\ref{fig:SM-Higgs}. 
The solid line depicts to the mass of the heavy CP-even Higgs boson ($m_H$) 
  while the dashed and dotted lines correspond to the CP-odd Higgs boson mass ($m_A$) 
  and the charged Higgs boson mass ($m_{H^\pm}$), respectively.

\section{Minimal extension with Higgs-portal dark matter}
If the SUSY breaking scale is 
$\sqrt{f}\lesssim 100$ TeV, gravitino mass is found to be $m_{3/2}\lesssim 10$ eV.
Although such a light gravitino (goldstino) is harmless in phenomenological point of view, 
  it is unable to be the dominant component of the DM in our universe and therefore 
  a suitable DM candidate should be supplemented to  the NL-MSSM.
In order to solve this problem, we propose a minimal extension of the NL-MSSM 
  to incorporate a dark matter candidate, namely, the (scalar) Higgs-portal DM.

The Higgs-portal DM scenario is one of the simplest SM extensions 
  to supplement the SM with a dark matter candidate. 
For a recent review, see Ref.~\cite{Arcadi:2019lka} and references therein.
In the simplest setup, we introduce an SM-singlet real scalar ($S$) along with a $Z_2$ symmetry. 
The stability of this scalar is ensured by assigning an odd-parity to it, 
  while all the SM fields are $Z_2$-even. 
At the renormalizable level, the Lagrangian is given by
\begin{eqnarray}
{\cal L}={\cal L}_{\rm SM}+{1 \over 2} (\partial_\mu S) (\partial^\mu S) 
  -{1 \over 2}M_S^2S^2-{1 \over 4}\lambda_SS^4- \frac{1}{4}\lambda_{HSS} (H^\dagger H) S^2,
\label{HP-int}
\end{eqnarray}
where ${\cal L}_{\rm SM}$ is the Lagrangian of the SM, and $H$ is the SM Higgs doublet field. 
After the electroweak symmetry breaking, the Lagrangian becomes
\begin{eqnarray}
{\cal L}={\cal L}_{\rm SM}+{1 \over 2} (\partial_\mu S)  (\partial^\mu S)
-{1 \over 2}m_{DM}^2 S^2-{1 \over 4}\lambda_SS^4- \frac{1}{4}\lambda_{HSS} \, v \, h \, S^2-{1 \over 8}\lambda_{HSS}h^2S^2,
\end{eqnarray}
where $h$ is the physical Higgs boson, and the DM mass $m_{DM}$ is given by
\begin{eqnarray}
 m_{DM}^2=M_S^2+{1 \over 4}\lambda_{HSS}v^2.
\end{eqnarray}
Here, the vacuum expectation value of the Higgs field is set to be $\langle H \rangle=(0,v)^T/\sqrt{2}$ with $v=246$ GeV.
The DM phenomenology in this Higgs-portal DM scenario is controlled by only two free parameters:  
  $m_{DM}$  and $\lambda_{HSS}$. 

\begin{figure}[tb]\label{Fig:cross}
\centering
\includegraphics[scale=0.2, angle=0]{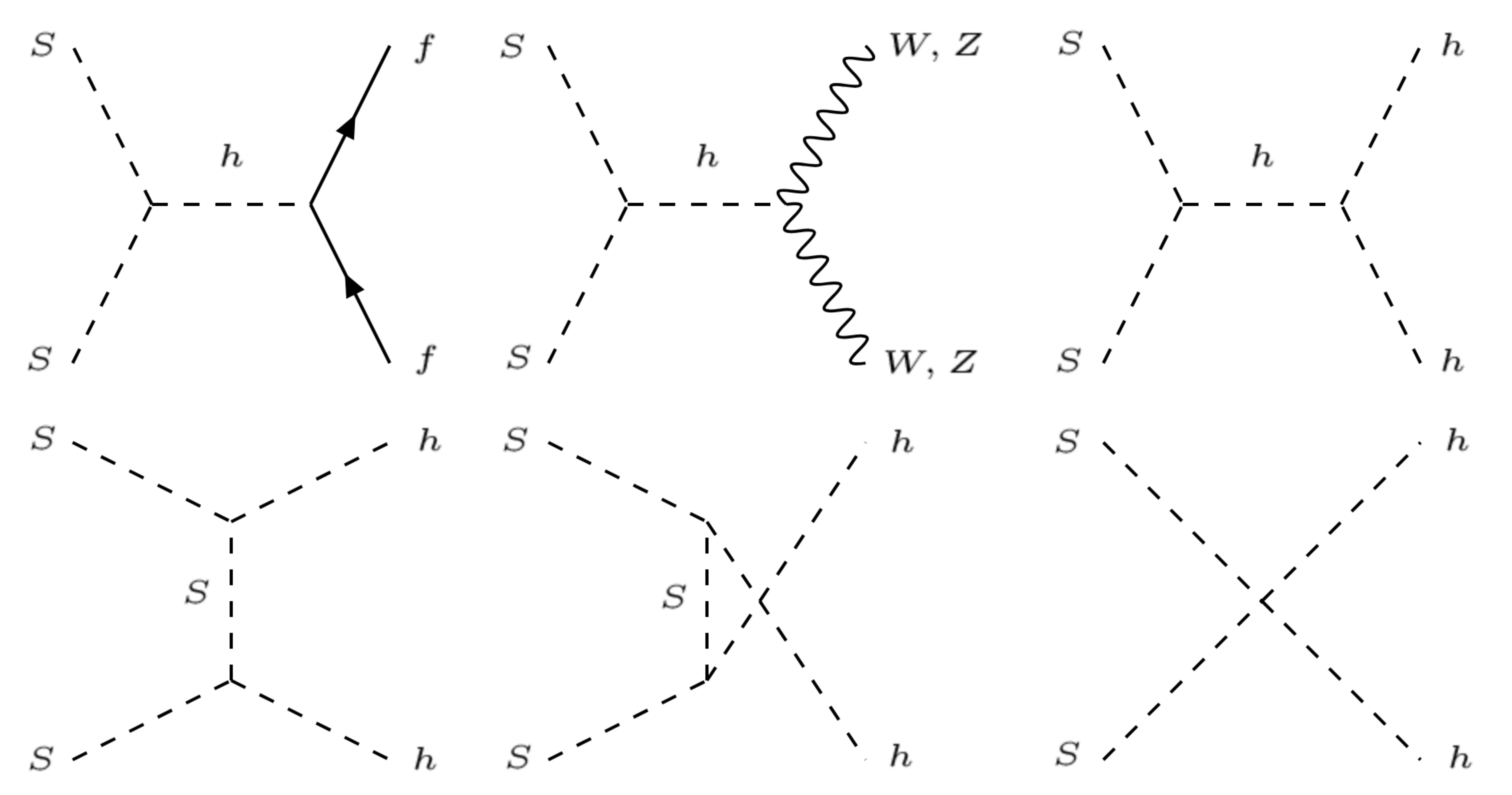}
 \caption{Feynman diagrams for dark matter annihilations.}
\end{figure}
The scalar DM $S$ annihilates into the SM particles through its coupling with the Higgs boson. 
The annihilation processes are shown in Fig.~\ref{Fig:cross}, 
   where $W(Z)$ is the charged (neutral) weak gauge boson, and $f$ represents quarks and leptons in the SM.
 We evaluate the DM relic density by solving the Boltzmann equation \cite{Gondolo:1990dk}:
\begin{eqnarray}
\frac{d Y}{d x}
= -\frac{s(m_{DM})}{H(m_{DM})} \,  \frac{\langle\sigma v_{\rm rel} \rangle}{x^2} \, (Y^2-Y_{EQ}^2) ,
 \label{eq:BE}
\end{eqnarray}
where the temperature of the universe is normalized by the DM mass as $x=m_{DM}/T$, 
   $H(m_{DM})$ and  $s(m_{DM})$ are the Hubble parameter,  
   and the entropy density of the universe at $T=m_{DM}$, respectively, 
   $Y= n/s$ is the DM yield (the ratio of the DM number density ($n$) to the entropy density ($s$)), 
   $Y_{EQ}$ is the yield of the DM particle in thermal equilibrium, 
  and $\langle \sigma v_{\rm rel} \rangle$ is the thermal-averaged DM annihilation cross section 
  times relative velocity ($v_{\rm rel}$). 
The formulas for the quantities in the Boltzmann equation are given as follows: 
\begin{eqnarray} 
s(T)= \frac{2  \pi^2}{45} g_\star T^3 ,  \; \;
H(T) =  \sqrt{\frac{\pi^2}{90} g_\star} \frac{T^2}{M_P},  \; \;
n_{EQ}=s \, Y_{EQ}= \frac{g_{DM}}{2 \pi^2} \frac{m_{DM}^3}{x} K_2(x),   
\end{eqnarray}
where $M_P=2.43 \times 10^{18}$  GeV is the reduced Planck mass, 
   $g_{DM}=1$ is the number of degrees of freedom for the Higgs-portal DM, 
   $g_\star$ is the effective number of total degrees of freedom for the particles in thermal equilibrium 
   ($g_\star=106.75$ for the SM particles),  
   and $K_2$ is the modified Bessel function of the second kind.   
The thermal-averaged annihilation cross section is calculated by
\begin{eqnarray}
\langle \sigma v \rangle = \frac{1}{n_{EQ}} \, g_{DM}^2 \,
  \frac{m_{DM}}{64 \pi^4 x} 
  \int_{4 m_{DM}^2}^\infty  ds \,  
  2 (s- 4 m_{DM}^2) \, \sigma(s)  \, \sqrt{s} K_1 \left(\frac{x \sqrt{s}}{m_{DM}}\right) , 
\label{ThAvgSigma}
\end{eqnarray}
where $\sigma(s)$ is the DM pair annihilation cross section corresponding to the processes in Fig.~\ref{Fig:cross}, 
  and $K_1$ is the modified Bessel function of the first kind. 
By using the asymptotic value of the yield $Y(\infty)$, 
  the DM relic density $\Omega_{\rm DM}h^2$ is expressed by
\begin{eqnarray}
  \Omega_{\rm DM}h^2={m_{\rm DM}s_0 Y(\infty) \over \rho_c/h^2}, \label{eq:CPDM}
\end{eqnarray}
   where $s_0=2890~{\rm cm}^{-3}$ is the entropy density of the present universe, 
   and $\rho_c/h^2=1.05\times 10^{-5}~{\rm GeV}~{\rm cm}^{-3}$ is the critical density.

The resultant DM relic density is controlled by two free parameters ($m_{DM}$ and $\lambda_{HSS}$), 
  and their relation is determined so as to reproduce the observed DM relic density 
  of $ \Omega_{\rm DM}h^2=0.12$ \cite{Aghanim:2018eyx}. 
In addition to the DM relic density, the parameter space of $m_{DM}$ and $\lambda_{HSS}$ 
  are constrained by the direct/indirect DM particle search results and the Higgs-portal DM search results 
  by the Large Hadron Collider (LHC) experiment. 
After all the constraints are taken into account, the allowed parameter region is identified. 
For the result, see, for example, Fig.~19 in Ref.~\cite{Arcadi:2019lka}. 
It has been found that the Higgs-portal DM scenario is phenomenologically viable, 
  but the allowed parameter region is very limited:
  $m_{DM} \simeq M_h/2$
  with the SM Higgs boson mass 
 $M_h=125$ GeV 
  and $10^{-4}\lesssim |\lambda_{HSS}| \lesssim 10^{-3}$.

Now we introduce an SM-singlet chiral superfield $\Phi$ along with a $Z_2$ symmetry and 
  assign odd-parity to it while even-parity to all the MSSM fields. 
Hence, the lightest component field in $\Phi$ is stable and the DM candidate.   
The SUSY Lagrangian ${\cal L}_0$ in Eq.~(\ref{eq:SUSY_L}) is then extended to be
\begin{eqnarray}
{\cal L}_0 \to {\cal L}_0 + \int d^4\theta \, \Phi^\dagger \Phi +
\left\{ \int d^2 \theta \, \mu_\Phi \Phi^2 + {\rm h.c.} \right\} \,,
\label{eq:SUSY_Phi}
\end{eqnarray}
where $\mu_\Phi$ is a mass parameter.
Similar to ${\cal L}_H$ and ${\cal L}_m$, a new Lagrangian for $\Phi$ involving the goldstino chiral superfield is given by 
\begin{eqnarray}
{\cal L}_\Phi =- {m_\Phi^2 \over f^2} \int d^4\theta \left( X^\dagger X \right) \Phi^\dagger \Phi \,,
\label{eq:soft1_Phi}
\end{eqnarray}
where $m_\Phi$ denotes a soft SUSY breaking mass.
Finally, ${\cal L}_{AB}$ is extended to be
\begin{eqnarray}
{\cal L}_{AB} \to {\cal L}_{AB} +    \left\{- {B_\Phi \over 2f} \int d^2\theta \, X \Phi^2+{\rm h.c.} \right\} \, . 
\label{eq:soft2_Phi}
\end{eqnarray}
In the following, we assume that $B_\Phi$ is real and positive. 

We now read off the scalar potential relevant to the Higgs-portal DM scenario 
   by eliminating the auxiliary fields: 
\begin{eqnarray}
V=V_{\rm SUSY} + V_{\rm soft}\,,
\end{eqnarray}
where 
\begin{eqnarray}
 V_{\rm SUSY}&=&
 \mu_H^2(|H_u|^2+|H_d|^2)+\mu_\Phi^2|\Phi|^2
 +{g_Z^2 \over 8}(|H_u|^2-|H_d|^2)^2+{g_2^2 \over 2}|H_u^\dagger H_d|^2 \,, \\
 V_{\rm soft}&=&{\left|f+{m_3^2 \over f}H_uH_d- {B_\Phi \over 2f}\Phi^2\right|^2 \over 1-{m_1^2 \over f^2}|H_d|^2-{m_2^2 \over f^2}|H_u|^2-{m_{\Phi}^2 \over f^2}|\Phi|^2}\,.
\end{eqnarray}
Although the complete form of the scalar potential includes all the sfermions in the MSSM, 
  we have considered the potential terms involving only the MSSM Higgs doublets and the SM-singlet scalar $\Phi$.  
This is because the sfermions should be heavy to satisfy the current LHC constraints 
 and their couplings with the Higgs-portal DM have little effects on the DM physics
 for $m_{DM} \simeq M_h/2$.  
For the physics of the Higgs-portal DM scenario, only the bilinear terms with respect to $\Phi$ are important. 
To extract them from the scalar potential, we expand $V_{\rm soft}$ up to the order of ${\cal O}(1/f^2)$ and then obtain
\begin{eqnarray}
V & \supset &
\left[ \left( \mu_\Phi^2  + m_{\Phi}^2 \right) +
\left\{\left({m_3^2 \over f^2}H_uH_d+ {\rm h.c.}  \right)+2{m_1^2 \over f^2}|H_d|^2+2{m_2^2 \over f^2}|H_u|^2\right\}m_{\Phi}^2 \right]|\Phi|^2 \nonumber \\
 &&-\left\{\left(1+{m_3^2 \over f^2}H_uH_d+{m_1^2 \over f^2}|H_d|^2+{m_2^2 \over f^2}|H_u|^2\right){B_\Phi \over 2}\Phi^2+{\rm h.c.}\right\}
 \,. \label{eq:int}
\end{eqnarray}

Substituting 
\begin{eqnarray}
\Phi&=&{1 \over \sqrt{2}}(\phi+i \eta) 
\end{eqnarray}
into Eq.~(\ref{eq:int}), we can find the mass spectrum of the real scalars, $\phi$ and $\eta$, 
  and their couplings with the Higgs bosons. 
First, we obtain the mass spectrum to be 
\begin{eqnarray}
 m_{\phi/\eta}^2&=&
\mu_\Phi^2 + m_\Phi^2 
+ \left( 
{m_1^2 \over f} \cos^2\beta + 
{m_2^2 \over f} \sin^2\beta + 
{m_3^2  \over f} \sin\beta \cos\beta 
\right) {m_\Phi^2 \over f} v^2  \nonumber \\
&&
\mp 
\left\{
 1+  \left( 
{m_1^2 \over f} \cos^2\beta + 
{m_2^2 \over f} \sin^2\beta + 
{m_3^2  \over f} \sin\beta \cos\beta 
\right) {v^2 \over f}
\right\} B_\Phi   \nonumber \\ 
& \simeq &
\mu_\Phi^2 + m_\Phi^2  \mp B_\Phi. 
\label{DM-Mass}
 \end{eqnarray}
In the last expression, we have used $|m_{1,2,3}^2|, f \gg v^2$ and $m_\Phi^2 < f$ from the theoretical consistency.  
We see that $m_\phi < m_\eta$ and thus the real scalar $\phi$ is the DM candidate.

Since all the Higgs bosons except for the SM-like Higgs boson are heavy, 
   the DM physics is mainly controlled by the coupling of $\phi$ with the SM-like Higgs boson. 
For a large $\tan \beta$ value, such as $\tan \beta=10$ as we have used in Figs.~\ref{fig:SM-Higgs} and \ref{fig:Heavy-Higgs}, 
   the up-type Higgs doublet is approximately identified as the SM-like Higgs doublet. 
By employing this approximation  $H_u \simeq H$, 
   we can easily extract the coupling of $\phi$ with the SM-like Higgs doublet from Eq.~(\ref{eq:int})
   such that
\begin{eqnarray}
 {\cal L}_{int} \simeq  - {m_2^2 \over f^2} \left( m_\Phi^2- \frac{B_\Phi}{2} \right) (H^\dagger H) \phi^2 \, .
\end{eqnarray}
This is the formula to be compared with Eq.~(\ref{HP-int}) with the identification of $S=\phi$. 
Therefore, in the decoupling limit of the heavy Higgs bosons and all the MSSM sparticles, 
   we have obtained the Higgs-portal DM scenario as the low energy effective theory. 
In terms of our model parameters, the two parameters $m_{DM}=m_\phi$ and $\lambda_{HSS}$, 
   which control the Higgs-portal DM physics, are approximately expressed as    
\begin{eqnarray}
   m_{DM}^2 &\simeq&  \mu_\phi^2+ m_\Phi^2 -B_\Phi  \, , \nonumber \\
   \lambda_{HSS} &\simeq&  4 \, {m_2^2 \over f^2} \left( m_\Phi^2- \frac{B_\Phi}{2} \right) \,. \label{eq:para}
\end{eqnarray}
As in Fig.~\ref{fig:SM-Higgs}, 
  we may fix $m_2^2=- (2005)^2$ GeV$^2$ and $\sqrt{f} \geq 3990$ GeV
  so as to yield $m_h \leq125$ GeV at the tree-level.   
Even after this choice, we still have three free parameters, $\mu_\Phi$, $m_\Phi$ and $B_\Phi$
   and we can arrange them to satisfy the phenomenological constraints, 
   $m_{DM} \simeq M_h/2$ and $10^{-4}  \lesssim |\lambda_{HSS}| \lesssim 10^{-3}$
   for the Higgs-portal DM scenario.\footnote{
For a parameter choice to predict $m_h < 125$ GeV at the tree-level,  
  $M_h=125$ GeV should be reproduced by $M_h^2 =m_h^2 + \Delta m_h^2$ 
   with $\Delta m_h^2$ from quantum corrections through scalar top quarks, as usual in the MSSM.}  
For example, we may set $\mu_\Phi^2 \simeq m_\Phi^2 \simeq B_\Phi/2 = {\cal O}(1 \, {\rm TeV}^2)$ 
   but tune their differences so as to reproduce the allowed values of $m_{DM}^2 \ll 1 \, {\rm TeV}^2$ and $ |\lambda_{HSS}| \ll1$.

%
%
%
\section{Conclusion}

If SUSY is broken at a low energy, the NL-MSSM with the goldstino chiral superfield 
  is a very useful description for taking the hidden sector effect into account to the MSSM. 
The NL-MSSM may be particularly interesting if the SUSY breaking scale lies in the multi-TeV range. 
In this case the SM-like Higgs boson mass $m_h=125$ GeV is achieved 
  by the Higgs potential at the tree-level after eliminating the $F$-component of the goldstino superfield. 
However, such a low scale SUSY breaking predicts a milli-eV gravitino LSP, 
  which is too light to be the main component of the DM in our universe. 
Thus, a suitable DM candidate is missing in the NL-MSSM.   
To solve this problem, we have proposed a minimal extension of the NL-MSSM 
  by introducing the SM-singlet chiral superfield ($\Phi$) along with the $Z_2$ symmetry. 
The stability of the lightest component field in $\Phi$ is ensured 
  by assigning odd-parity to $\Phi$ while even-parity for all the MSSM superfields. 
We have shown that in the decoupling limit of the sparticles and heavy Higgs bosons, 
  our low energy effective theory is nothing but the Higgs-portal DM scenario 
  with the lightest $Z_2$-odd real scalar being the DM candidate. 
With a suitable choice of the model parameters, 
  we can reproduce the allowed parameter region of the Higgs-portal DM scenario.

Here we give a comment on a general property of our model. 
Since the main point of this paper is to propose the minimal extension 
  of the NL-MSSM to incorporate a suitable DM candidate, 
  we have focused on a special parameter region,
for which our model at low energies is reduced 
  to the simplest Higgs-portal DM scenario (plus a extremely light gravitino),  
  namely, the SM with the real scalar DM being odd under the $Z_2$ symmetry.
In general, we have a wide variety of the parameter choices 
  to realize a viable dark matter scenario.
For example, we may take a very small value of $B_\Phi$ in Eq.~(\ref{DM-Mass}) so that the mass splitting 
  between $\phi$ and $\eta$ is negligibly small. 
 In this case, we identify the complex scalar $\Phi$ with the DM particle. 
This is a complex scalar extension of the simplest Higgs-portal DM scenario with only one real scalar. 
This extension has been studied in \cite{McDonald:1993ex, Barger:2010yn,  Gonderinger:2012rd, Chiang:2012qa, Coimbra:2013qq, Costa:2015llh, Wu:2016mbe}.
Since the MSSM includes two Higgs doublets, our Higgs-portal DM scenario is basically 
  two Higgs doublet extension of the Higgs-portal DM scenario.
The two Higgs doublet extension of the SM supplemented by the Higgs-portal DM 
 has been studied in \cite{Bird:2006jd, He:2007tt, He:2008qm, Grzadkowski:2009iz, Aoki:2009pf, Li:2011ja, Cai:2011kb, He:2011gc, Bai:2012nv, He:2013suk, Greljo:2013wja, Wang:2014elb, Drozd:2014yla, Okada:2014usa, Campbell:2015fra, Drozd:2015gda}. 
In the model,
 the heavy Higgs bosons can play an important role for the DM physics, 
  for example, an enhancement of the DM pair annihilations through the heavy Higgs boson resonances. 
While the allowed parameter region of the simplest Higgs-portal DM scenario is very limited, 
  a wide parameter space can be phenomenologically viable in the two-Higgs doublet extension.
Although we focused on the interaction between the DM particle and the Higgs boson, 
  the full Lagrangian includes interactions between the DM particles and sfermions, 
  which are also derived by integrating out the $F$-component of the goldstino superfield. 
If  the DM particle is heavier than sfermions, the DM pair annihilation processes 
  through the interaction between the DM particle and sfermions become important in evaluating the DM relic density.   
We leave such general analysis for future work.

Finally, let us consider a crucial difference of our model from standard neutralino dark matter scenario in the MSSM.
Neutralino dark matter is an $R$-parity odd particle, while in our scenario the dark matter is an $R$-parity even particle.
This fact leads to distinctive phenomenologies. For example, in collider phenomenology, a neutralino dark matter is produced 
through a cascade decay of heavier sparticles due to the $R$-parity conservation. On the other hand, Higgs-portal
dark matter in our scenario is not produced by sparticle decays. A pair of Higgs-portal dark matters 
can be produced from the decays of Higgs bosons. Since gravitino is an $R$-parity odd particle, it is produced from 
a cascade decay of sparticles. In our scenario, the gravitino is almost massless, and missing energy distributions associated 
with gravitino production are very different from those associated with neutralino production.

%
%
\subsection*{Acknowledgments} 
N.O. would like to thank the High Energy Theory Group in Yamagata University 
  for the hospitality during his visit. 
This work is supported in part by the United States Department of Energy Grant No.~DE-SC0012447 (N.O.). 

%
%

\end{document}